# Use of Electron Paramagnetic Resonance (EPR) technique to build quantum computers: n-qubit (n=1,2,3,4) Toffoli Gates

Sayan Manna* and Sushil K. Misra**

Physics Department, Concordia University,

7420 Sherbrooke St. West, Montreal, Quebec H4B 1R6, Canada

**Abstract.** It is shown theoretically how to use the EPR (Electron Paramagnetic Resonance) technique, using electron spins as qubits, coupled with each other by the exchange interaction, to set the configuration of n qubits (n=1,2,3,4) at resonance, in conjunction with pulses, to construct the NOT (one qubit), CNOT (two qubits), CCNOT (three qubits), CCCNOT (four qubits) Toffoli gates, which can be exploited to build a quantum computer. This is unique to EPR, wherein exchange-coupled electron spins are used. This is difficult with NMR (Nuclear Magnetic Resonance), that uses nuclear spins as qubits, which do not couple with each other by the exchange interaction.

*MITACS summer intern (2024) from IIT Kharagpur, India.

** Corresponding author.

## 1. **Introduction**.

Logic gates are the fundamental building blocks of digital circuits. They are used to perform basic operations like AND, OR, and NOT. They can be used to make counters, storage selectors, adders, and more. One can test logic gates by using a truth table, which shows all the outputs for all possible input combinations. In order to build a computer system, one can use logic gates that process binary data (1s and 0s). These are: AND gate, which outputs a 1 if both inputs are 1; OR gate, which outputs a 1 if either input is 1; NOT gate (also known as an inverter), which outputs the opposite value of its input; NAND gate:, which returns false only if both inputs are true; NOR gate, which outputs true only when both inputs are false; XOR gate, which returns true when the inputs vary; XNOR gate (also known as NOT-XOR), which produces a true output when the inputs are equal.

In classical computers, it is inherently impossible to construct logical gates. One of the impediments is the irreversibility of logical gates, i.e., those of AND & OR gates. This restricts the computational speed of the classical computer. Here comes the quantum computers [1] and quantum gates. There are several approaches for constructing logical gates. One of them was proposed by Feynman [2]. He proved that every logical gate can be constructed by combining only two reversible gates: a one-qubit (quantum bit) gate e.g., NOT, Hadamard and the other with two qubit Controlled-NOT (CNOT) gate. Besides, the papers in Refs. [3, 4, 5] have showed high-fidelity CNOT gates. To perform a fault-tolerant quantum computation [6, 7] (Appendix I), one can exploit a discrete set of gates, which are Feynman gates. In general, any unitary operation on n qubits can be described by composing single qubit and CNOT gates.

There are several approaches for realizing quantum logical gates. One of these is to exploit nuclear magnetic resonance (NMR). However, it is difficult to make a logical gate by NMR for several qubits, because the inter-spin interactions among the nuclei are quite weak, not amenable to the approach proposed here. On the other hand, EPR [8, 9] (Electron Paramagnetic Resonance, also known as ESR – Electron Spin Resonance), is quite suitable for this purpose, since there, indeed, exist reasonably strong, required inter-electron exchange interaction, not possible amongst nuclei.

It is the purpose of this paper to exploit EPR to construct the following gates: NOT (one qubit), CNOT (two qubits), CCNOT (three qubits) [10], and CCCNOT (four qubits). These gates are also known as Toffoli gates. Quantum gates based on EPR [11] are performed by subjecting the electron spins to a microwave electromagnetic field controlling them under the action of an external magnetic field (Zeeman field) [12]. By exploiting EPR, Ohya, Volovich and Watanabe [13] worked out the details of constructing a NOT-like gate using uncoupled single electron spins, and by additionally invoking the Ising model of exchange interactions between two electron spins, they extended this approach to construct the CNOT gates based on EPR. Subsequently, Miyuzami, Watanabe, Volovich [14] used this approach to extend this treatment to three qubits i.e., to construct the CCNOT gate, based on EPR. This paper proposes an EPR-based approach to construct the three-qubit CCNOT gate, different from that proposed in [14], which is further extended here to the four-qubit CCCNOT gate based on EPR.

The advantages of a four-qubit CCCNOT gate are as follows. The CCNOT gate, which is the standard Toffoli gate with two control qubits, and the CCCNOT gate, with three control qubits, are both multi-controlled quantum gates, which are used in quantum computing. The primary difference between them lies in the number of control qubits they have. Some advantages of the CCCNOT gate over the CCNOT gate are as follows. *(i) Increased Control and Selectivity.* Because of three control qubits, the CCCNOT gate allows for more selective operation, meaning the target qubit will only flip if all three control qubits are in the state $|1\rangle$. This provides a higher degree of control to the CCCNOT gate, as compared to the CCNOT gate, which only requires two control qubits to be in the state $|1\rangle$. This increased control is advantageous in complex quantum algorithms, where more precise conditional operations are needed. *(ii) Use in Complex Quantum Algorithms.* The CCCNOT gate finds its better use in algorithms that require more layers of conditional logic. For example, additional control qubits help, in reversible computing or certain error-correcting codes, to implement more sophisticated logic without requiring extra ancillary qubits or gates. In Grover's algorithm and other quantum search algorithms, gates with more control qubits, like CCCNOT, can help create more efficient oracle implementations for specific problem sets. *(iii) Reduction in Ancillary Qubits.* One can sometimes reduce the number of ancillary (helper) qubits needed by directly using a CCCNOT gate instead of building an equivalent operation from multiple CCNOT gates. In this manner, one can reduce the overall qubit count and circuit complexity. *(iv) Simplification of Circuit Design.* By using a CCCNOT gate one can simplify the overall circuit design in certain quantum circuits. A single CCCNOT gate might implement the same operation more straightforwardly, instead of cascading multiple CCNOT gates, which could introduce additional intermediate qubits and complexity. *(v) Implementation of Higher-Order Multi-Control Gates.* The CCCNOT gate serves as a basis for building even more complex quantum gates, associated with more control qubits. One can build circuits with even higher degrees of control (e.g., CCCCCCNOT) by understanding and implementing CCCNOT gates.

The organization of this paper is as follows. For completeness, details of constructing all four - one qubit (NOT), two qubit (CNOT), three qubit (CCNOT), and four-qubit (CCCNOT) gates are included here. The general procedure to construct EPR-based Toffoli gates are described in Sec. 2. Specific details of construction of n-qubit (n = 1,2,3,4) Toffoli gates are given in Sec. 3. Conclusions are summarized in Sec. 4.

## 2. General procedure to construct an n-qubit EPR-based Toffoli gate

First, a unitary operator $U_{\phi}$ , based on EPR at resonance, is derived in the *rotating frame* (Appendix II) for the system of n qubits. This is further augmented by pulses needed in the desired Toffoli gate for the corresponding quantum circuit, as necessary. To this end, the system of qubits is subjected to EPR, assuming the qubits (n > 1) to be electron spins coupled by the exchange interaction. Theoretically, one treats the system quantum mechanically in the rotating frame of reference, as described in Appendix II, creating a $U_{\phi}(t)$ operator for the resonance condition: $B_0 = \frac{\omega}{\gamma}$ , where $B_0$ is the external Zeeman field and $\omega$ is the frequency of the applied microwave field, with $\gamma$ (=$1.76085963 \times 10^{11} rad \cdot s^{-1} \cdot T^{-1}$) being the *gyromagnetic ratio* of the electron. Thereafter, the required pulses are applied to point the spins in the desired orientations to accomplish the various Toffoli gates. For the systems of multiple qubits, the details of constructing the respective Toffoli gate are as follows.

## 3. Details of construction of n-qubit (n = 1,2,3,4) gates

### 3.1 NOT gate based on EPR (n=1)

Here a particular value of time $t_1$ is chosen such that the $U_{\phi}(t_1)$ required is converted to the NOT gate by a subsequent application of a pulse applied during the time $t_2$. The mathematical details of accomplishing the NOT gate are as follows.

Let the Hilbert space $\mathcal{H}$ be $\mathbb{C}^2$ , where $\mathbb{C}^2$ is the two-dimensional complex vector space. The basis states are:

$$u_+ = |\uparrow\rangle = \begin{bmatrix}1\\0\end{bmatrix}$$

$$u_- = |\downarrow\rangle = \begin{bmatrix}0\\1\end{bmatrix}$$

Here $u_+$ and $u_-$ are the spin vectors corresponding to the spin up and spin down states of the qubit, respectively.

Let, $\vec{S} = (S_x, S_y, S_z)$ be a spin angular momentum operator of electron, where $S_i = \frac{\sigma_i}{2}$ is a component of spin operator of electron in direction $i$ $(i = x, y, z)$, were $\sigma_i$ is a Pauli spin operator. The unit vectors along the $x, y, z$ axes are denoted as $\overrightarrow{e_x}, \overrightarrow{e_y}, \overrightarrow{e_z}$ .

Then,

$$\vec{S} = (S_x, S_y, S_z) = S_x\overrightarrow{e_x} + S_y\overrightarrow{e_y} + S_z\overrightarrow{e_z}$$

Let us consider two magnetic fields $\overrightarrow{B_0}$ and $\overrightarrow{B_1}$. $\overrightarrow{B_0}$ is the static magnetic field in the $z$ direction given by $\overrightarrow{B_0} = B_0\overrightarrow{e_z}$, whereas $\overrightarrow{B_1}$ is the rotating magnetic field with frequency ω in the $xy$ plane, given by

$$\overrightarrow{B_1}(t) = B_1(\overrightarrow{e_x}\cos(\omega t) - \overrightarrow{e_y}\sin(\omega t))$$

where, $B_0$ and $B_1$ are constant amplitudes of the static and rotating magnetic fields, respectively. So, the total magnetic field is:

$$\vec{B}(t) = \vec{B}_0 + \overrightarrow{B_1}(t)$$

The corresponding Hamiltonian is.

$$H_s = -\gamma\,\vec{S}\cdot\vec{B} = -\gamma[B_0S_z + B_1(S_x\cos(\omega t) - S_y\sin(\omega t))]$$

Consider the initial state to be $\psi(0) = a_0|\uparrow\rangle + b_0|\downarrow\rangle = \begin{bmatrix}a_0\\b_0\end{bmatrix}$ .

Then, the state vector at time $t$ is denoted by

$$\psi(t) = a(t)|\uparrow\rangle + b(t)|\downarrow\rangle = \begin{bmatrix} a(t) \\ b(t) \end{bmatrix}$$

where $a(t), b(t) \epsilon \mathbb{C}$ and satisfy the normalizing condition $|a(t)|^2 + |b(t)|^2 = 1$

Now the time dependent Schrödinger equation is

$$i\hbar \frac{\partial}{\partial t}\psi(t) = \widehat{H_s}\psi(t) = -\gamma[B_0 S_z + B_1(S_x \cos(\omega t) - S_y \sin(\omega t))]\psi(t),$$

where $\widehat{H_s}$ is the spin Hamiltonian of the one-qubit system, i.e., that of one electron.

The solution to this equation, as shown in Appendix II, is given in the rotating frame, by

$$\psi(t) = e^{i\omega t S_z} e^{it\gamma((B_0 - \frac{\omega}{\gamma})S_z + B_1 S_x)}\psi(0)$$

The resonance condition is: $\omega = B_0\gamma$. Then, the state at resonance is:

$$\psi(t) = e^{it\gamma B_0 S_z} e^{it\gamma B_1 S_x}\psi(0)$$

Now, if we take $t = t_1$ such that $\frac{\gamma B_0 t_1}{2} = 2n\pi + \frac{\pi}{2}$ and $\frac{\gamma B_1 t_1}{2} = 2m\pi + \frac{\pi}{2}$, where $n, m$ are integers and $n >> m$. (The second of the above condition can be achieved by adjusting the value of $B_1$ once the first condition is adjusted by choosing $t_1$.). Then,

$$\psi(t_1) = e^{i(2n\pi + \frac{\pi}{2})\sigma_z} e^{i(2m\pi + \frac{\pi}{2})\sigma_x}\psi(0)$$

Since, $e^{iAx} = \cos(x) I + i \sin(x) A$, where $A$ is an operator, one obtains

$$\psi(t_1) = e^{i\frac{\pi}{2}\sigma_z} e^{i\frac{\pi}{2}\sigma_x}\psi(0) = i^2 \sigma_z \sigma_x \psi(0)$$

Therefore,

$$\psi(t_1) = \begin{bmatrix} 0 & -1 \\ 1 & 0 \end{bmatrix}\psi(0)$$

Now, to create the NOT gate, one needs to apply a pulse to cause a rotation of $\psi(t_1)$ by the angle $-\frac{\pi}{2}$ around the $z$-axis, for which the required pulse is $e^{i\frac{\pi}{2}\sigma_z}$, which takes the time interval $t_2 = \frac{\pi}{2\omega}$ to apply. If this pulse is applied to $\psi(t_1)$, one obtains:

$$\psi(t_1 + t_2) = e^{i\frac{\pi}{2}\sigma_z}\begin{bmatrix} 0 & -1 \\ 1 & 0 \end{bmatrix}\psi(0) = -i\begin{bmatrix} 0 & 1 \\ 1 & 0 \end{bmatrix}\psi(0)$$

So, the NOT gate, $\begin{bmatrix} 0 & 1 \\ 1 & 0 \end{bmatrix}$, is obtained here with a phase factor of $-i$.

### 3.2 Two-qubit CNOT gate based on EPR (n=2)

As there are two qubits here, the coupling exchange interaction comes into play. First, by solving the Schrödinger equation in the rotating frame of reference and applying the resonance condition, one obtains $U_{\phi}\,(t)$, the time evolution operator, according to which the joint state vector of two qubits evolves, in the same manner as that used above for one qubit.

To accomplish the CNOT gate, the following quantum circuit, given by Figure 1, is used.

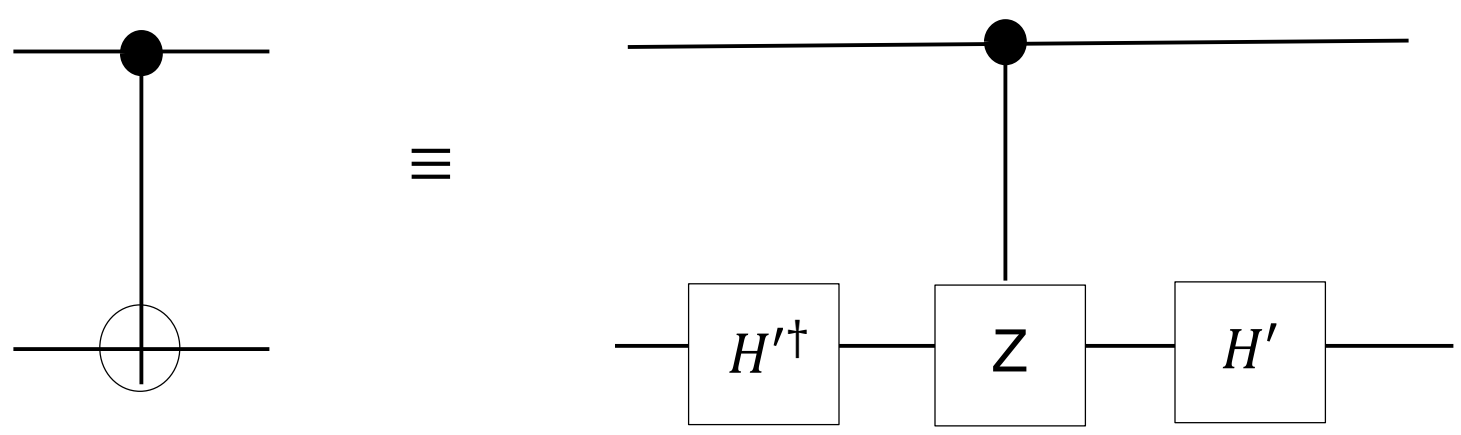

Figure 1: Quantum circuit to construct CNOT gate for two qubits

It is difficult to produce the exact Hadamard gate, H, using EPR, instead, a Hadamard-like gate, denoted by $H'$, is produced. In the above diagram, $H'^{\dagger}$ is the adjoint of $H'$, which is the Hadamard like gate, the middle gate is the Controlled-$Z$ gate. (details given below)

The theoretical details of accomplishing the CNOT gate are described as follows.

Let $u_+, u_-, v_+, v_-$, be the basis spin vectors in the two- dimensional Hilbert space for two qubits; here $u$ refers to the first qubit and $v$ refers to the second qubit and +, - refer to the up and down spin orientations.

The basis states of 2 qubits are, in the direct product space,

$$u_+ \otimes v_+ = \begin{bmatrix} 1 \\ 0 \\ 0 \\ 0 \end{bmatrix} \quad u_+ \otimes v_- = \begin{bmatrix} 0 \\ 1 \\ 0 \\ 0 \end{bmatrix} \quad u_- \otimes v_+ = \begin{bmatrix} 0 \\ 0 \\ 1 \\ 0 \end{bmatrix} \quad u_- \otimes v_- = \begin{bmatrix} 0 \\ 0 \\ 0 \\ 1 \end{bmatrix},$$

where $\otimes$ stands for the direct product

Let $\psi(0)$ be the initial state vector given by,

$$\psi(0) = a_0 u_+ \otimes v_+ + b_0 u_+ \otimes v_- + c_0 u_- \otimes v_+ + d_0 u_- \otimes v_- = \begin{bmatrix} a_0 \\ b_0 \\ c_0 \\ d_0 \end{bmatrix},$$

where, $a_0, b_0, c_0, d_0 \in \mathbb{C}^4$, where $\mathbb{C}^4$ is four-dimensional complex vector space.

The total magnetic field is the sum of the static field ($\overrightarrow{B_0}$) along the $z$-axis and the rotating field ($\overrightarrow{B_1}$), in the $xy$ plane, as follows:

$$\vec{B}(t) = B_0 \hat{e}_z + B_1(\hat{e}_x \cos(\omega t) - \hat{e}_y \sin(\omega t))$$

where $\hat{e}_x, \hat{e}_y, \hat{e}_z$ are the unit vectors along the $x, y, z$ axes. For the two qubits, the two spin vectors are denoted as $\overrightarrow{S_1}$, $\overrightarrow{S_2}$ :

$$\overrightarrow{S_1} = S_{x1}\hat{e}_x + S_{y1}\hat{e}_y + S_{z1}\hat{e}_z$$

$$\overrightarrow{S_2} = S_{x2}\hat{e}_x + S_{y2}\hat{e}_y + S_{z2}\hat{e}_z$$

The corresponding Hamiltonian for the two qubits is

$$H = -\gamma(\overrightarrow{S_1} + \overrightarrow{S_2}) \cdot \vec{B} + J(\overrightarrow{S_1} \cdot \overrightarrow{S_2})$$

where $J$ is the exchange coupling constant between the two electrons.

Thus, the total Hamiltonian is

$$H = -\gamma[B_0(S_{z1} + S_{z2}) + B_1(\cos(\omega t)(S_{x1} + S_{x2}) - \sin(\omega t)(S_{y1} + S_{y2}))]$$
$$+J(S_{x1} \cdot S_{x2} + S_{y1} \cdot S_{y2} + S_{z1} \cdot S_{z2})$$

Assuming now the exchange coupling interaction term of the Hamiltonian to be of the Ising type, $J(S_{z1} \cdot S_{z2})$ ; $J > 0$. Then, in the rotating frame of reference, the time-independent Hamiltonian is

$$H_{(2)} = -\gamma[(B_0 - \frac{\omega}{\gamma})(S_{z1} + S_{z2}) + B_1(S_{x1} + S_{x2})] + J(S_{z1} \cdot S_{z2})$$

To create the CNOT gate for two qubits, the reference energy is changed by adding the constant energy term $B'I$, where $I$ is the 4x4 identity matrix. Then, the two-qubit Hamiltonian becomes

$$H'_{(2)} = -\gamma[(B_0 - \frac{\omega}{\gamma})(S_{z1} + S_{z2}) + B_1(S_{x1} + S_{x2})] + J(S_{z1} \cdot S_{z2}) + B'I$$

Now the solution to the Schrödinger equation with the new Hamiltonian $H'_{(2)}$ in the rotating frame at resonance ($B_0 = \frac{\omega}{\gamma}$) is given by

$$\psi(t) = exp\,[\frac{i\omega t(S_{z1}+S_{z2})}{\hbar}]\; exp\,[-\frac{iH'_{(2)}t}{\hbar}]\psi(0)$$

(See Appendix II). For simplification, changing now the energy unit so that $\hbar = 1$, one obtains

$$\psi(t) = e^{i\omega t(S_{z1}+S_{z2})}e^{-it[-\gamma B_1(S_{x1}+S_{x2})+J(S_{z1}\cdot S_{z2})+B']}\psi(0),$$

which can be expressed symbolically as

$$\psi(t) = U_\phi(t)\psi(0)$$

where $U_\phi(t)$ is the time-evolution unitary operator.

(N.B, the following notation, in the direct-product space of the two qubits is used hereafter:
$S_{z1} = S_z \otimes I$ ; $S_{z2} = I \otimes S_z$ ; $S_{x1} = S_x \otimes I$ ; $S_{x2} = I \otimes S_x$ ; $S_{z1} \cdot S_{z2} = S_z \otimes S_z$)

So, the time-evolution operator for two qubits, at resonance, is (rotating frame)

$$U_\phi(t) = e^{i\gamma B_0 t(S_z\otimes I)}e^{i\gamma B_0 t(I\otimes S_z)}e^{i\gamma\ {}_1t(S_x\otimes I)}e^{i\gamma B_1 t(I\otimes S_x)}e^{-iJt(S_z\otimes S_z)}e^{-iB't(I\otimes I)} \qquad (1)$$

The details of the two additional component gates needed to construct the Toffoli gate, are described as follows.

### 3.2.1 Controlled-Z gate

To build the CNOT gate we need to create the Controlled-Z gate, where the first qubit is the control qubit, and the second qubit is the target qubit. So, only when the first qubit is in state 1 will the Z gate be applied to the second qubit.

The Controlled-Z gate will now be created from $U_\phi$. To create the Controlled-Z gate, the value of $B_1$ and $t = t_1$ are chosen, such that

$$\gamma B_0 t_1 = \omega t_1 = 2n\pi + \frac{\pi}{2} \quad \text{and } \gamma B_1 t_1 = 2m\pi$$

where $m, n$ are integers, such that $n > 0\ and\ m > 0, n >> m$

For $U_\phi$ to become the Controlled-Z matrix, one needs to choose $B'$ and $J$, such that the following condition are simultaneously satisfied:

$$Jt_1 = (2p+1)\pi,$$

$$B't_1 = (2q + \tfrac{1}{4})\pi, \qquad (2)$$

where $p, q$ are integers.

Then, using the above-mentioned values, one obtains from Equation (1):

$$U_\phi(t_1) = e^{i(2n\pi+\frac{\pi}{2})(S_z\otimes I)} e^{i(2n\ +\frac{\pi}{2})(I\otimes S_z)} e^{i(2m\pi)(S_x\otimes I)} e^{i(2m\pi)(I\otimes S_x)} e^{-i(2p\pi+\pi)(S_z\otimes S_z)} e^{-i(2q\pi+\frac{\pi}{4})(I\otimes I)}$$

$$= e^{i(\frac{\pi}{2})(S_z\otimes I)} e^{i(\frac{\pi}{2})(I\otimes S_z)} II e^{-i(\pi)(S_z\otimes S_z)} e^{-i(\frac{\pi}{4})(I\otimes I)}$$

$$= e^{i(\frac{\pi}{4})(\sigma_z\otimes I)} e^{i(\frac{\pi}{4})(I\otimes \sigma_z)} II e^{-i(\frac{\pi}{4})(\sigma_z\otimes \sigma_z)} e^{-i(\frac{\pi}{4})(I\otimes I)}$$

$$\Rightarrow U_\phi(t_1) = \begin{bmatrix} 1 & 0 & 0 & 0 \\ 0 & 1 & 0 & 0 \\ 0 & 0 & 1 & 0 \\ 0 & 0 & 0 & -1 \end{bmatrix} \tag{3}$$

The above matrix, given by Equation (3), is the matrix for the desired Controlled-$Z$ gate.

### 3.2.2 Hadamard-like gate ($U_{H'}$)

Next, a unitary operator $U_{H'}(t)$, related to Hadamard transformation, based on EPR, is created. This transformation can be achieved by applying a pulse to the second qubit, rotating it around the $y$-axis and leaving the first qubit intact.

Thus, $U_{H'}(t)$is defined as:

$$U_{H'}(t) = e^{-i\omega t(I\otimes S_y)}$$

where $H'$ is the required Hadamard-like operator, acting only on the second qubit, is expressed in the matrix form as follows:

$$H' = \frac{1}{\sqrt{2}}\begin{bmatrix} 1 & -1 \\ 1 & 1 \end{bmatrix}.$$

Now, following the quantum circuit, as given in Figure 1, starting with the initial state vector $\psi(0)$, after the application of the first $H'^{\dagger}$ gate over the time interval $t_2$, one obtains

$$\psi(t_2) = e^{-i\omega t_2(I\otimes S_y)}\psi(0) = U_{H'}\psi(0)$$

This unitary operator, $U_{H'}$ , does a Hadamard-like transformation on the second qubit and does nothing to the first qubit. So, the corresponding matrix turns out to be as follows:

$$U_{H'}(t_2), = \begin{bmatrix} 1 & 0 \\ 0 & 1 \end{bmatrix} \otimes \frac{1}{\sqrt{2}}\begin{bmatrix} 1 & -1 \\ 1 & 1 \end{bmatrix}$$

To achieve $U_{H'}(t_2)$, one chooses $t_2$ for the Hadamard-like pulse such that $\omega t_2 = \frac{\pi}{2}$.

Thus, the unitary operator $U_{CNOT}$ related to the CNOT gate can be constructed by the combination of $U_{\phi}(t_1)$, which is the control-Z gate with the first qubit as the control qubit and the second qubit as the target qubit, and $U_{H'}(t_2)$ (Hadamard-like), as shown below in Figure 2.

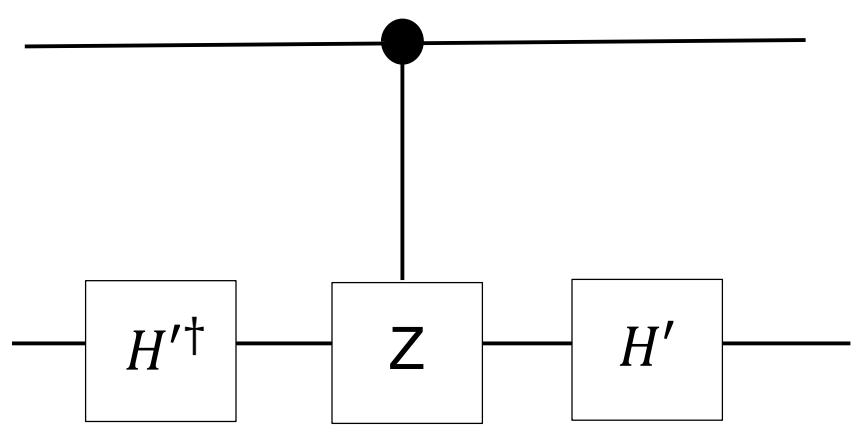

Figure 2: Quantum circuit for CNOT gate

Accordingly, first a pulse relevant to producing $U^{\dagger}{}_{H'}$ to the second qubit is applied over the time duration of $t_2$, thereafter the resultant state of the two qubits is subjected to EPR to apply the operator $U_{\phi}(t_1)$ (Controlled-$Z$) over the time duration $t_1$, and thereafter a pulse $U_{H'}$ is applied to the second qubit to set it to the desired quantum state. This total operation is equivalent to the CNOT gate, accomplished over the total time of $t_1 + 2t_2$, shown as argument of $U_{CNOT}(t_1 + 2t_2)$, to get the desired quantum state.

Therefore,

$$U_{CNOT}(t_1 + 2t_2) = U_{H'}(t_2)U_{\phi}(t_1)U^{\dagger}{}_{H'}(t_2)$$

$$\Rightarrow U_{CNOT}(t_1 + 2t_2) = e^{-i\omega t_2 t_2 (I \otimes S_y)} e^{i\omega t_1 (S_z \otimes I)} e^{i\omega t_1 (I \otimes S_z)} e^{-iJt_1 (S_z \otimes S_z)} e^{-iB' t_1 (I \otimes I)} e^{i\omega_2 t_2 (I \otimes S_y)} \tag{4}$$

Now, one chooses $t_2$ such that $\omega t_2 = \frac{\pi}{2}$, and by substituting the values of $\omega t_1$, $Jt_1$, $B't_1$ in Equation (4), as given in Equation (2), one obtains

$$U_{CNOT}(t_1 + 2t_2) = \begin{bmatrix} 1 & 0 & 0 & 0 \\ 0 & 1 & 0 & 0 \\ 0 & 0 & 0 & 1 \\ 0 & 0 & 1 & 0 \end{bmatrix}$$

This is the desired CNOT matrix, corresponding to the two-qubit Toffoli gate.

### 3.3 Three-qubit CCNOT gate based on EPR (n=3)

This gate will be created according to the following quantum circuit (Figure 3) [15]:

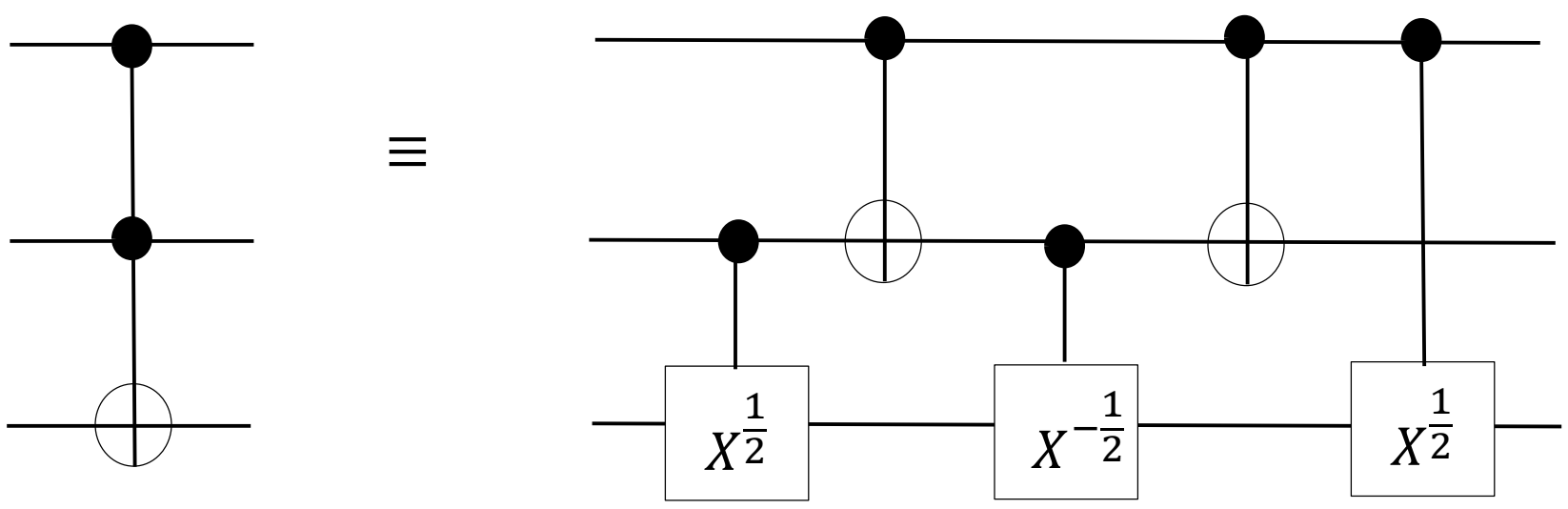

Figure 3: Quantum circuit to construct CCNOT gate for three qubits

The details for constructing the CNOT gate are as follows.

Here, one has three qubits (electron spins), for which, there will be needed three pairwise exchange interactions, $J_{12}, J_{23}, J_{13}$. By arranging these three spins in space such that they are at the vertices of an equilateral triangle, the values of all the pairwise exchange interaction become the same, i.e., $J_{12} = J_{23} = J_{13} = J$.

The Hamiltonian for these spins will be, then

$$H = -\gamma(\overrightarrow{S_1}+\overrightarrow{S_2} + \overrightarrow{S_3}) \cdot \vec{B} + J[\overrightarrow{S_1} \cdot \overrightarrow{S_2} + \overrightarrow{S_2} \cdot \overrightarrow{S_3} + \overrightarrow{S_1} \cdot \overrightarrow{S_3}]$$

where, the first term is the Zeeman term, and the last three terms are exchange-coupling terms. Assuming now that the exchange interaction is of Ising-type, and the external static magnet field, $B_0$,is oriented in $z$-direction, one obtains

$$H = -\gamma[B_0(S_{z1} + S_{z2} + S_{z3}) + B_1[cos\,(\omega t)(S_{x1} + S_{x2} + S_{x3})$$

$$- sin\,(\omega t)(S_{y1} + S_{y2} + S_{y3})\,] \; + J[S_{z1} \cdot S_{z2} + S_{z2} \cdot S_{z3} + S_{z1} \cdot S_{z3}]$$

The above is expressed in the rotating frame of reference, where the frame is rotating about the $z$-axis with the angular frequency $\omega$. In this frame of reference, the total Hamiltonian is

$$H_{(3)} = -\gamma\left[\left(B_0 - \frac{\omega}{\gamma}\right)(S_{z1} + S_{z2} + S_{z3}) + B_1(S_{x1} + S_{x2} + S_{x3})\right]$$

$$+J[S_{z1} \cdot S_{z2} + S_{z2} \cdot S_{z3} + S_{z1} \cdot S_{z3}]$$

To create the gate, one needs to change the reference energy by adding the constant energy term $B'I$, where $I$ is the 8x8 identity matrix. Then

$$H'_{(3)} = -\gamma\left[\left(B_0 - \frac{\omega}{\gamma}\right)(S_{z1} + S_{z2} + S_{z3}) + B_1(S_{x1} + S_{x2} + S_{x3})\right]$$

$$+\mathrm{J}[S_{z1} \cdot S_{z2} + S_{z2} \cdot S_{z3} + S_{z1} \cdot S_{z3}] + B'I$$

The solution to the Schrödinger equation, in the rotating frame of reference at resonance, is given by

$$\psi(t) = exp\,[i\omega t(S_{z1} + S_{z2} + S_{z3})]\; exp\,[-iH'_{(3)}t]\psi(0)$$

(See Appendix II; here the energy is expressed in units of $\hbar$.)

Then the time-dependent solution is

$$\psi(t) = e^{iB_0\gamma t(S_{z1}+S_{z2}+S_{z3})}e^{i\gamma t B_1(S_{x1}+S_{x2}+S_{x3})}e^{-it[J(S_{z1}\cdot S_{z2}+S_{z2}\cdot S_{z3}+S_{z1}\cdot S_{z3})+B']}\psi(0),$$

where $\psi(0)$ is the wave function at time t=0.

The corresponding time evolution operator, $U_\phi(t)$, is expressed at resonance as

$$U_\phi(t) = e^{iB_0\gamma t(S_{z1}+S_{z2}+S_{z3})}e^{i\gamma t B_1(S_{x1}+S_{x2}+S_{x3})}e^{-it\mathrm{J}[S_{z1}\cdot S_{z2}+S_{z2}\cdot S_{z3}+S_{z1}\cdot S_{z3}]}e^{-itB'}$$

since $\omega = B_0\gamma$ at resonance. Now, if one chooses the duration of time t, such that $\frac{B_1\gamma t}{2} = 2m\pi$, where m > 0. is an integer. Then, the operator containing the $S_x$ terms in $U_\phi$ will be just the Identity operator as $e^{i\frac{\gamma t B_1}{2}(\sigma_{x1}+\sigma_{x2}+\sigma_{x3})} = I$.

Thus,

$$U_\phi(t) = e^{i\omega t(S_z\otimes I \otimes I)}e^{i\omega t(I\otimes S_z\otimes I)}e^{i\omega t(I\otimes I \otimes S_z)}e^{-iJt(S_z\otimes S_z \otimes I)}$$

$$\times\; e^{-iJt(S_z\otimes I \otimes S_z)}e^{-iJt(I\otimes S_z \otimes S_z)}e^{-iB't(I\otimes I \otimes I)}$$

$$\Rightarrow U_\phi(t) = e^{i\frac{\omega t}{2}(\sigma_z\otimes I \otimes I)}e^{i\frac{\omega t}{2}(I\otimes\sigma_z\otimes I)}e^{i\frac{\omega t}{2}(I\otimes I \otimes\sigma_z)}e^{-i\frac{Jt}{4}(\sigma_z\otimes \sigma_z \otimes I)}$$

$$\times\, e^{-i\frac{Jt}{4}(\sigma_z\otimes I\otimes\sigma_z)}e^{-i\frac{Jt}{4}(I\otimes \sigma_z \otimes\sigma_z)}e^{-iB't(I\otimes I \otimes I)} \quad (5)$$

***The above Equation** (5) **for** $U_\phi$ **is the key equation***. This $U_\phi(t)$ will be used hereafter to create the various component gates only by changing the values of $\omega t, Jt, B't$.

Now, in the quantum circuit shown in Figure 3, the component gates from left to right are:

- **Controlled-$X^{\frac{1}{2}}$** (controlled by qubit 2 with the target on qubit 3), denoted as ${U^{(2,3)}}_{SX}$ (where the notation used is such that ${U^{(2,3)}}_{SX}$ stands for the qubits 2 and 3 connected by the Controlled-$X^{\frac{1}{2}}$ gate.
- **CNOT** (controlled by qubit 1 and the target on qubit 2), denoted as ${U^{(1,2)}}_{CNOT}$; ${U^{(1,2)}}_{CNOT}$ stands for the qubits 1 and 2 connected by the CNOT gate.
- **Controlled-$X^{-\frac{1}{2}}$** (controlled by qubit 2 and the target on qubit 3), denoted as $({U^{(2,3)}}_{SX})^\dagger$;
- **CNOT** (controlled by qubit 1 and the target on qubit 2), denoted as ${U^{(1,2)}}_{CNOT}$;
- **Controlled-$X^{\frac{1}{2}}$** (controlled by qubit 1 and the target on qubit 3), denoted as ${U^{(1,3)}}_{SX}$,

The various component gates will now be derived theoretically by applying selective pulses on $U_\phi$, as follows.

**Timings for different quantum component gates**

**Controlled-$X^{\frac{1}{2}}$ gate:**

To obtain it, one needs to exploit $U_\phi$, as given by Equation (5). Choose now the time periods $t_j$ ($j$ = 1 & 6), for gates ${U^{(2,3)}}_{SX}$ and ${U^{(1,3)}}_{SX}$, respectively, such that

$$\frac{\omega t_j}{2} = 2n\pi - \frac{\pi}{8};\ n > 0$$

$$\frac{Jt_j}{4} = 2m\pi - \frac{\pi}{8};\ m > 0$$

$$B't_j = 2p\pi - \frac{\pi}{8};\ p > 0$$

where *n, m, p* are integers. One obtains from Equation (5):

$$U_\phi(t_j) = e^{-i\frac{\pi}{8}(\sigma_{z1}+\sigma_{z2}+\sigma_{z3})} e^{i\frac{\pi}{8}(\sigma_{z1}\cdot\sigma_{z2}+\sigma_{z1}\cdot\sigma_{z3}+\sigma_{z2}\cdot\sigma_{z3})} e^{i\frac{\pi}{8}(I\otimes I\otimes I)}$$

($j$ = 1 & 6)

The timings will be different for different gates. They are defined below.

- For the controlled-$X^{\frac{1}{2}}$ gate, with the second qubit as the control qubit and the third qubit as the target qubit (denoted by ${U^{(2,3)}}_{SX}$), $j = 1$. The related $U_\phi$ gate is denoted as $U_\phi(t_1)$.

- For the controlled-$X^{\frac{1}{2}}$ with the first qubit as the control qubit and the third qubit as the target qubit (denoted by $U^{(1,3)}{}_{SX}$), $j = 6$. The related $U_\phi$ gate is denoted as $U_\phi(t_6)$.

Controlled-$X^{\frac{1}{2}}$ gate for three qubits, where the second qubit is the control qubit and the third qubit is the target qubit, and the first qubit is inactive, is shown below.

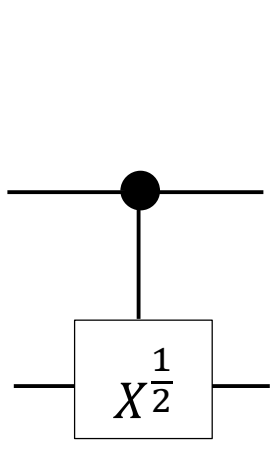

It is expressed as

$$U^{(2,3)}{}_{SX}(t_1 + 2t_2 + 3t_3) = e^{-i\omega t_2 S_{y3}} U_\phi(t_1) e^{2i\omega t_3 (S_{z1} \cdot S_{z2})} e^{2i\omega t_3 (S_{z1} \cdot S_{z3})} e^{-i\omega t_3 S_{z1}} e^{i\omega t_2 S_{y3}}$$

where the argument $t_1 + 2t_2 + 3t_3$ of $U^{(2,3)}{}_{SX}$ on the left-hand side is the total time taken by the applied pulses.

If one now chooses the following values for the durations $t_2, t_3$, such that

$$\frac{\omega t_2}{2} = 2m\pi + \frac{\pi}{4};\ m > 0$$

$$\frac{\omega t_3}{2} = 2n\pi - \frac{\pi}{8};\ n > 0$$

where $m, n$ are integers. Then,

$$U^{(2,3)}{}_{SX}(t_1 + 2t_2 + 3t_3) = e^{-i\omega t_2 S_{y3}} U_\phi(t_1) e^{2i\omega t_3 (S_{z1} \cdot S_{z2})} e^{2i\omega t_3 (S_{z1} \cdot S_{z3})} e^{-i\omega t_3 S_{z1}} e^{i\omega t_2 S_{y3}}$$

$$\Rightarrow U^{(2,3)}{}_{SX}(t_1 + 2t_2 + 3t_3) = e^{-i\frac{\pi}{4}\sigma_{y3}} U_\phi(t_1) e^{-i\frac{\pi}{8}(\sigma_{z1} \cdot \sigma_{z2})} e^{-i\frac{\pi}{8}(\sigma_{z1} \cdot \sigma_{z3})} e^{i\frac{\pi}{8}\sigma_{z1}} e^{i\frac{\pi}{4}\sigma_{y3}}$$

**CNOT gate:**

The CNOT ($U^{(1,2)}{}_{CNOT}$) quantum circuit for 3 qubits, where (1,2) indicates that the first qubit is the control qubit, and the second qubit is the target qubit, to which the NOT gate will be applied, and the third qubit is inactive, is as follows:

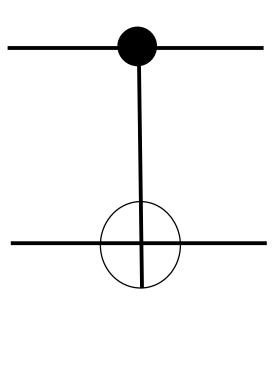

To create this CNOT gate, one chooses the duration $t = t_4$ in $U_\phi(t)$ , given by Equation (4), such that

$$\frac{\omega t_4}{2} = 2n\pi + \frac{\pi}{4};\ n > 0$$
$$\frac{Jt_4}{4} = 2m\pi + \frac{\pi}{4};\ m > 0$$
$$B' t_4 = 2p\pi + \frac{\pi}{4};\ p > 0$$

where $m, n, p$ are integers.

Then,

$$U_\phi(t_4) = e^{i\frac{\pi}{4}(\sigma_{z1}+\sigma_{z2}+\sigma_{z3})} e^{-i\frac{\pi}{4}(\sigma_{z1}\cdot\sigma_{z2}+\sigma_{z1}\cdot\sigma_{z3}+\sigma_{z2}\cdot\sigma_{z3})} e^{-i\frac{\pi}{4}(I\otimes I\otimes I)}$$

Thus,

$$U^{(1,2)}{}_{CNOT}(T) = e^{-i\omega t S_{y2}} U_\phi(t_4) e^{2i\omega t(S_{z1}\cdot S_{z3})} e^{2i\omega t(S_{z2}\cdot S_{z3})} e^{-i\omega t S_{z3}} e^{i\omega t S_{y2}}$$

where T is the total time taken by the applications of the various pulses.

The value of $t_5$ is now chosen, such that

$$\frac{\omega t_5}{2} = 2m\pi + \frac{\pi}{4};\ m > 0$$

where, $m$ is an integer. Then, taking account all the above-defined timings, one obtains

$$U^{(1,2)}{}_{CNOT}(t_4 + 5t_5) = e^{-i\omega t_5 S_{y2}} U_\phi(t_4) e^{2i\omega t_5(S_{z1}\cdot S_{z3})} e^{2i\omega t_5(S_{z2}\cdot S_{z3})} e^{-i\omega t_5 S_{z3}} e^{i\omega t_5 S_{y2}}$$

$$= e^{-i\frac{\pi}{4}\sigma_{y2}} U_\phi(t_4) e^{i\frac{\pi}{4}(\sigma_{z1}\cdot\sigma_{z3})} e^{i\frac{\pi}{4}(\sigma_{z2}\cdot\sigma_{z3})} e^{-i\frac{\pi}{4}\sigma_{z3}} e^{i\frac{\pi}{4}\sigma_{y2}}$$

$$= \begin{bmatrix} 1 & 0 & 0 & 0 \\ 0 & 1 & 0 & 0 \\ 0 & 0 & 0 & 1 \\ 0 & 0 & 1 & 0 \end{bmatrix} \otimes I$$

This is the CNOT gate where the first qubit is the control qubit, and the second qubit is the target qubit.

❖ *Controlled-$X^{-\frac{1}{2}}$ gate, $({U^{(2,3)}}_{SX})^\dagger$), for 3 qubits where the second qubit is the control qubit, and the third qubit is the target qubit; the first qubit is inactive. Its quantum circuit is shown as follows:*

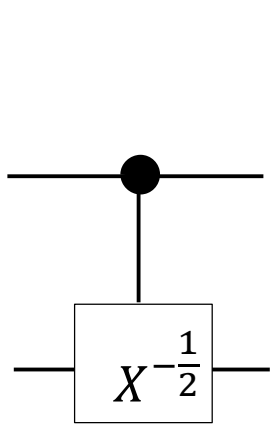

This gate is, in fact, $({U^{(2,3)}}_{SX})^\dagger$, i.e., the adjoint of ${U^{(2,3)}}_{SX}$, which has already been discussed above.

❖ *Controlled-$X^{\frac{1}{2}}$ gate, (${U^{(1,3)}}_{SX}$), for the for 3 qubits, where the first qubit is the control qubit and the third qubit is the target qubit, and the second qubit is inactive is shown as follows:*

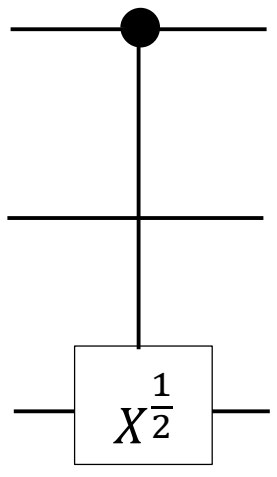

The operator for this circuit can be obtained similarly to that for $U^{(2,3)}{}_{SX}$ by interchanging the qubits 2 and 1 with each other, except for the matrix multiplication $(S_{z1} \cdot S_{z2})$, which remains the same. In addition, the pulse timings change from $t_1$ to $t_6$, $t_2$ to $t_7$, and $t_3$ to $t_8$.
Then,

$$U^{(1,3)}{}_{SX}(t_6 + 2t_7 + 3t_8) = e^{-i\frac{\pi}{4}\sigma_{y3}} U_{\phi}(t_6) e^{-i\frac{\pi}{8}(\sigma_{z1}\cdot\sigma_{z2})} e^{-i\frac{\pi}{8}(\sigma_{z2}\cdot\sigma_{z3})} e^{i\frac{\pi}{8}\sigma_{z2}} e^{i\frac{\pi}{4}\sigma_{y3}}$$

## Construction of CCNOT gate

Now, the component gates from left to right corresponding to the quantum circuit, as shown in Figure 3, are:

$U^{(2,3)}{}_{SX}$ , $U^{(1,2)}{}_{CNOT}$, $(U^{(2,3)}{}_{SX})^{\dagger}$, $U^{(1,2)}{}_{CNOT}$, $U^{(1,3)}{}_{SX}$

denoting now

$$T_1 = t_1 + 2t_2 + 3t_3$$
$$T_2 = t_4 + 5t_5$$
$$T_3 = t_6 + 2t_7 + 3t_8$$
$$T = 2T_1 + 2T_2 + T_3$$

the three-qubit $8 \times 8$ Toffoli matrix is obtained by multiplying together the various component gates, shown in Figure 3:

$$U_{CCNOT}(T) = U^{(1,3)}{}_{SX}(T_3) \cdot U^{(1,2)}{}_{CNOT}(T_2) \cdot (U^{(2,3)}{}_{SX})^{\dagger}(T_1) \cdot U^{(1,2)}{}_{CNOT}(T_2) \cdot U^{(2,3)}{}_{SX}(T_1)$$

$$= \begin{bmatrix} 1 & 0 & 0 & 0 & 0 & 0 & 0 & 0 \\ 0 & 1 & 0 & 0 & 0 & 0 & 0 & 0 \\ 0 & 0 & 1 & 0 & 0 & 0 & 0 & 0 \\ 0 & 0 & 0 & 1 & 0 & 0 & 0 & 0 \\ 0 & 0 & 0 & 0 & 1 & 0 & 0 & 0 \\ 0 & 0 & 0 & 0 & 0 & 1 & 0 & 0 \\ 0 & 0 & 0 & 0 & 0 & 0 & 0 & 1 \\ 0 & 0 & 0 & 0 & 0 & 0 & 1 & 0 \end{bmatrix},$$

which is the three-qubit $8 \times 8$ Toffoli matrix.

### 3.4 CCCNOT gate based on EPR (n=4)

This gate is created by using the following quantum circuit [15]:

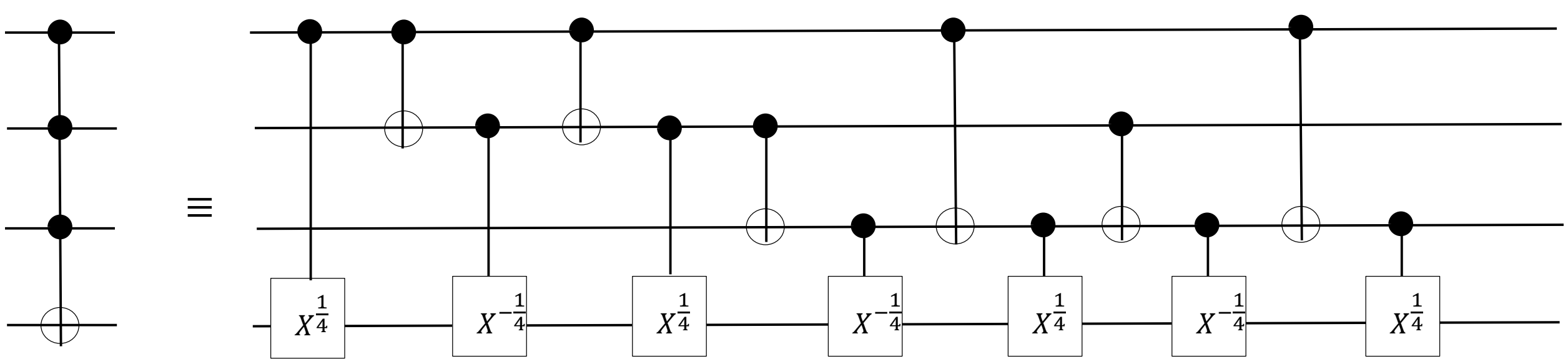

Figure 4: Quantum circuit for four-qubit CCCNOT gate

Here the required component gates are Controlled-$X^{\frac{1}{4}}$ $(4\ times)$, CNOT $(6\ times)$, Controlled-$X^{-\frac{1}{4}}$ $(3\ times)$.

Now, as there are four qubits, there will be required a total of six exchange-coupling interactions. To build the gate, it will simplify if one were to consider all the six pairwise interactions to be of the same magnitude, $J$. Also, one needs to include all six pairwise interaction terms, because to build this CCCNOT gate all of them are required. To achieve it, the four qubits must be situated at the vertices of a regular tetrahedron, as shown below in Figure 5.

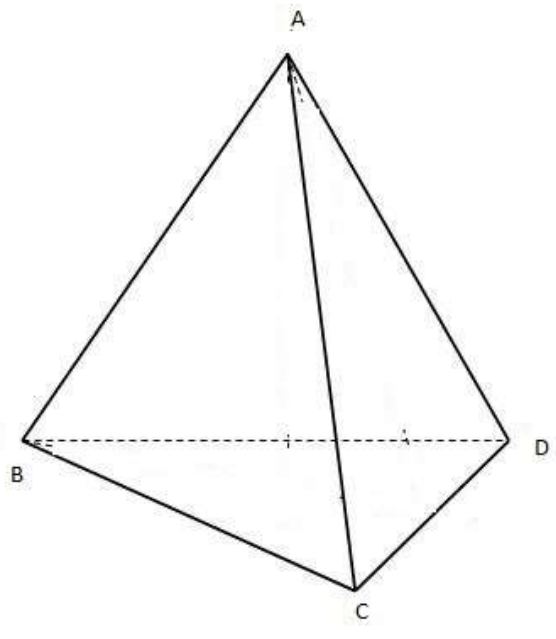

Figure 5: Configuration of four qubits on the vertices of a regular tetrahedron

The theoretical details of constructing the CCCNOT gate are as follows. The total magnetic field vector for EPR, consisting of a static external magnetic field in the $z$-direction and a rotating magnetic field in the $xy$ plane, is:

$$\vec{B}(t) = B_0 \hat{e}_z + B_1 \{\hat{e}_x \, cos\,(\omega t) - \hat{e}_y \, sin\,(\omega t)\}$$

The four spin operators for the qubits are denoted as $\vec{S_1}, \vec{S_2}, \vec{S_3}, \vec{S_4}$

The Hamiltonian for the four qubits, coupled by the exchange interaction, is then

$$H = -\gamma(\vec{S_1} + \vec{S_2} + \vec{S_3} + \vec{S_4}) \cdot \vec{B} + J(\vec{S_1} \cdot \vec{S_2} + \vec{S_1} \cdot \vec{S_3} + \vec{S_1} \cdot \vec{S_4} + \vec{S_2} \cdot \vec{S_3} + \vec{S_2} \cdot \vec{S_4} + \vec{S_3} \cdot \vec{S_4}),$$

where, the first term is the Zeeman interaction, and the second term is the coupling term consisting of six exchange-coupling interactions. Assuming now the exchange interaction to be of the Ising type,

$$\begin{gathered} H = -\gamma[B_0(S_{z1} + S_{z2} + S_{z3} + S_{z4}) \\ +B_1(\cos\,(\omega t)(S_{x1} + S_{x2} + S_{x3} + S_{x4}) \\ -\sin\,(\omega t)(S_{y1} + S_{y2} + S_{y3} + S_{y4})\,] \\ +J(S_{z1} \cdot S_{z2} + S_{z1} \cdot S_{z3} + S_{z1} \cdot S_{z4} + S_{z2} \cdot S_{z3} + S_{z2} \cdot S_{z4} + S_{z3} \cdot S_{z4}) \end{gathered}$$

In the frame of reference, rotating about the $z$-axis with the angular frequency, $\omega$, the time-independent Hamiltonian is

$$\begin{gathered} H_{(4)} = -\gamma[(B_0 - \frac{\omega}{\gamma})(S_{z1} + S_{z2} + S_{z3} + S_{z4}) + B_1(S_{x1} + S_{x2} + S_{x3} + S_{x4})] \\ +J(S_{z1} \cdot S_{z2} + S_{z1} \cdot S_{z3} + S_{z1} \cdot S_{z4} + S_{z2} \cdot S_{z3} + S_{z2} \cdot S_{z4} + S_{z3} \cdot S_{z4}) \end{gathered}$$

Now, to create the gate, the reference energy must be changed by adding a constant energy term $B'I$, where $I$ is the 16x16 identity matrix. Then,

$$\begin{gathered} H'_{(4)} = -\gamma[(B_0 - \frac{\omega}{\gamma})(S_{z1} + S_{z2} + S_{z3} + S_{z4}) + B_1(S_{x1} + S_{x2} + S_{x3} + S_{x4})] \\ +J(S_{z1} \cdot S_{z2} + S_{z1} \cdot S_{z3} + S_{z1} \cdot S_{z4} + S_{z2} \cdot S_{z3} + S_{z2} \cdot S_{z4} + S_{z3} \cdot S_{z4}) + B'I \end{gathered}$$

The solution to this Schrödinger equation in the rotating frame of reference at resonance is given by (Appendix II):

$$\psi(t) = exp\,[i\omega t(S_{z1} + S_{z2} + S_{z3} + S_{z4})]\; exp\,[-iH'_{(4)}t]\psi(0)$$

(expressed in unit of angular frequency of $\hbar = 1$)

At resonance, $\omega = B_0\gamma$, and by substituting for $H'_{(4)}$ one obtains,

$$\psi(t) = e^{i\omega t(S_{z1}+S_{z2}+S_{z3}+S_{z4})} e^{i\gamma t B_1(S_{x1}+S_{x2}+S_{x3}+S_{x4})}$$
$$\times\; e^{-i\;[J(S_{z1}\cdot S_{z2}+S_{z1}\cdot S_{z3}+S_{z1}\cdot S_{z4}+S_{z2}\cdot S_{z3}+S_{z2}\cdot S_{z4}+S_{z3}\cdot S_{z4})+B']}\psi(0)$$

This can be formally expressed as

$$\psi(t) = U_\phi(t)\psi(0)$$

Where (rotating frame),

$$U_\phi(t) = e^{i\omega t(S_{z1}+S_{z2}+S_{z3}+S_{z4})} e^{iB_1\gamma t(S_{x1}+S_{x2}+S_{x3}+S_{x4})} e^{-itJ(S_{z1}\cdot S_{z2}+S_{z1}\cdot S_{z3}+S_{z1}\cdot S_{z4}+S_{z2}\cdot S_{z3}+S_{z2}\cdot S_{z4}+S_{z3}\cdot S_{z4})} e^{-itB'}$$

Now, to create the gate, the $S_x$ term should be eliminated. This is accomplished by choosing the time duration $t$, such that

$B_1\gamma t = 2m\pi$, where $m$ is an integer > 0

Then, the operator containing the $S_x$ terms in $U_\phi$ will be just the Identity operator, since

$$e^{i\gamma t B_1(S_{x1}+S_{x2}+S_{x3}+S_{x4})} = I.$$

Then, using these

$$U_\phi(t) = e^{i\omega t(S_{z1}+S_{z2}+S_{z3}+S_{z4})} e^{-itJ(S_{z1}\cdot S_{z2}+S_{z1}\cdot S_{z3}+S_{z1}\cdot S_{z4}+S_{z2}\cdot S_{z3}+S_{z2}\cdot S_{z4}+S_{z3}\cdot S_{z4})} e^{-itB'} \qquad (6)$$

where, the following notations are used in the direct-product space: $S_{z1} = S_z \otimes I \otimes I \otimes I$; $S_{z2} = I \otimes S_z \otimes I \otimes I$ ; $S_{z3} = I \otimes I \otimes S_z \otimes I$; $S_{z4} = I \otimes I \otimes I \otimes S_z$; $S_{z1} \cdot S_{z2} = S_z \otimes S_z \otimes I \otimes I$; $S_{z1} \cdot S_{z3} = S_z \otimes I \otimes S_z \otimes I$; $S_{z1} \cdot S_{z4} = S_z \otimes I \otimes I \otimes S_z$; $S_{z2} \cdot S_{z3} = I \otimes S_z \otimes S_z \otimes I$; $S_{z2} \cdot S_{z4} = I \otimes S_z \otimes I \otimes S_z$; $S_{z3} \cdot S_{z4} = I \otimes I \otimes S_z \otimes S_z$.

***The above Equation (6) for $U_\phi$ is the key equation***. This $U_\phi$ will be used to create the various component gates to build the four qubit CCCNOT gate: Controlled-$X^{\frac{1}{4}}$, Controlled-$X^{-\frac{1}{4}}$, and CNOT gates.

It is noted that in the quantum circuit, given in Figure 4, the component gates are, in going from left to right, are as follows:

- Controlled-$X^{\frac{1}{4}}$ (controlled qubit one and target qubit four),
- CNOT (controlled qubit 1 and target qubit 2),
- Controlled-$X^{-\frac{1}{4}}$ (controlled qubit 2 and target qubit 4),
- CNOT (controlled qubit 1 and target qubit 2),
- Controlled-$X^{\frac{1}{4}}$ (controlled qubit 2 and target qubit 4),
- CNOT (controlled qubit 2 and target qubit 3),
- Controlled-$X^{-\frac{1}{4}}$ (controlled qubit 3 and target qubit 4),
- CNOT (controlled qubit 1 and target qubit 3),
- Controlled-$X^{\frac{1}{4}}$ (controlled qubit 3 and target qubit 4),
- CNOT (controlled qubit 2 and target qubit 3),
- Controlled-$X^{-\frac{1}{4}}$ (controlled qubit 3 and target qubit 4),
- CNOT (controlled qubit 1 and target qubit 3),
- Controlled-$X^{\frac{1}{4}}$ (controlled qubit 3 and target qubit 4).

All these component gates will be derived hereafter using the $U_\phi$ as given by Equation (6). This will be done by applying additional pulses to $U_\phi$. It is noted that there will be chosen different values of $\omega t, Jt, B't$ for different component gates. As there are different gates, their timings will be different, denoted by different $j$ indices on $t_j$. Below, the timings for the various gates are specified. It is noted that if it is the same gate between two same qubits, or its inverse, then the timings will be the same.

## Timings for different quantum component gates

**Controlled-$X^{\frac{1}{4}}$ gate:**

In $U_\phi$ given by Equation (6), choose a time interval $t = t_j$, *j = 1, 6, 11,* such that

$$\frac{\omega t_j}{2} = 2n\pi - \frac{\pi}{16};\ n > 0$$
$$\frac{Jt_j}{4} = 2m\pi - \frac{\pi}{16};\ m > 0$$
$$B' t_j = 2p\pi - \frac{\pi}{16};\ p > 0$$

one obtains,

$$U_\phi(t_j) = e^{-i\frac{\pi}{16}(\sigma_{z1}+\sigma_{z2}+\sigma_{z3}+\sigma_{z4})} e^{i\frac{\pi}{16}(\sigma_{z1}\cdot\sigma_{z2}+\sigma_{z1}\cdot\sigma_{z3}+\sigma_{z1}\cdot\sigma_{z4}+\sigma_{z2}\cdot\sigma_{z3}+\sigma_{z2}\cdot\sigma_{z4}+\sigma_{z3}\cdot\sigma_{z4})} e^{i\frac{\pi}{16}(I\otimes I\otimes I\otimes I)}$$

*(j = 1, 6, 11)*

(7)

For these gates, as far as the choice of the index $j$ is concerned, it is noted that

- For the Controlled-$X^{\frac{1}{4}}$ gate with the first qubit as the control qubit and the fourth qubit as the target qubit (denoted by $U^{(1,4)}{}_{SSX}$), one chooses $j = 1$ , so this gate will contain $U_\phi(t_1)$ in its expression.
- For the Controlled-$X^{\frac{1}{4}}$ gate with the second qubit as the control qubit and the fourth qubit as the target qubit (denoted by $U^{(2,4)}{}_{SSX}$), one chooses $j = 6$ , so this gate will contain $U_\phi(t_6)$ in its expression.
- For the Controlled-$X^{\frac{1}{4}}$ gate with the third qubit as the control qubit and the fourth qubit as the target qubit (denoted by $U^{(3,4)}{}_{SSX}$), one chooses $j = 11$ , so this gate will contain $U_\phi(t_{11})$ in its expression.

**CNOT gate:**

In $U_\phi$ given by Equation (6), choose a time interval $t = t_k$, *k = 4, 9, 14,* such that

$$\frac{\omega t_k}{2} = 2n\pi + \frac{\pi}{4};\ n > 0$$
$$\frac{Jt_k}{4} = 2m\pi + \frac{\pi}{4};\ m > 0$$
$$B' t_k = 2p\pi + \frac{\pi}{4};\ p > 0$$

where, $m, n, p$ are integers

Then, for these values of $t_k$ one obtains

$$U_{\phi}{}^{CNOT}(t_k) =$$

$$e^{i\frac{\pi}{4}(\sigma_{z1}+\sigma_{z2}+\sigma_{z3}+\sigma_{z4})}e^{-i\frac{\pi}{4}(\sigma_{z1}\cdot\sigma_{z2}+\sigma_{z1}\cdot\sigma_{z3}+\sigma_{z1}\cdot\sigma_{z4}+\sigma_{z2}\cdot\sigma_{z3}+\sigma_{z2}\cdot\sigma_{z4}+\sigma_{z3}\cdot\sigma_{z4})}e^{-i\frac{\pi}{4}(I\otimes I\otimes I\otimes I)} \quad (8)$$

For different CNOT gates with different control qubits and different target qubits, the timings will be different. They are defined as follows.

- For the CNOT gate, with the first qubit as the control qubit and the second qubit as the target qubit (denoted by $U^{(1,2)}{}_{CNOT}$), one chooses $k = 4$, so this gate will contain $U_{\phi}{}^{CNOT}(t_4)$ in it.
- For the CNOT gate, with the second qubit as the control qubit and the third qubit as the target qubit (denoted by $U^{(2,3)}{}_{CNOT}$), one chooses $k = 9$, so this gate will contain $U_{\phi}{}^{CNOT}(t_9)$ in it.
- For the CNOT gate, with the first qubit as the control qubit and the third qubit as the target qubit (denoted by $U^{(1,2)}{}_{CNOT}$), one chooses $k = 14$, so this gate will contain $U_{\phi}{}^{CNOT}(t_{14})$ in it.

Now, the formulation of the various gates is described as follows.

❖ *Controlled-$X^{\frac{1}{4}}$ gate, ($U^{(1,4)}{}_{SSX}$), for four qubits where the first qubit is the control qubit and the fourth qubit is the target qubit, second and the third qubits are inactive, is shown below.*

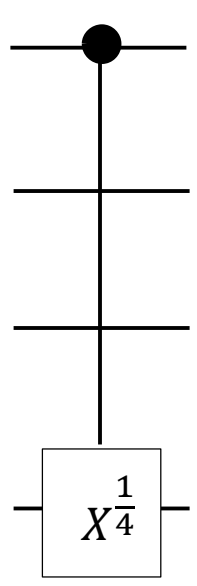

This gate will here be denoted by $U^{(1,4)}{}_{SSX}$, where (1,4) indicates that the first qubit is the control qubit, and the fourth qubit is the target qubit. Then

$$U^{(1,4)}{}_{SSX}(T) = e^{-i\omega t_2 S_{y4}}U_{\phi}(t_1)e^{2i\omega t_3(S_{z1}\cdot S_{z2})}e^{2i\omega t_3(S_{z1}\cdot S_{z3})}e^{2i\omega t_3(S_{z2}\cdot S_{z3})}e^{2i\omega t_3(S_{z2}\cdot S_{z4})}e^{2i\omega t_3(S_{z3}\cdot S_{z4})}$$

$$\times e^{-i\omega t_3 S_{z2}}e^{-i\omega t_3 S_{z3}}e^{i\omega t_2 S_{y4}}$$

where T is the total time taken by the pulses.

It is noted that $U_\phi(t_1)$ has already been defined above by Equation (7). The values of $t_2, t_3$ are now chosen such that

$$\frac{\omega t_2}{2} = 2m\pi + \frac{\pi}{4};\ m > 0,$$

$$\frac{\omega t_3}{2} = 2n\pi - \frac{\pi}{16};\ n > 0;$$

where, $m, n$ are integers.

Then, one obtains

$$U^{(1,4)}{}_{SSX}(t_1 + 2\,t_2 + 7\,t_3) = e^{-i\omega t_2 S_{y4}} U_\phi(t_1) e^{2i\omega t_3 (S_{z1}\cdot S_{z2})} e^{2i\omega t_3 (S_{z1}\cdot S_{z3})} e^{2i\omega t_3 (S_{z2}\cdot S_{z3})}$$

$$\times e^{2i\omega t_3 (S_{z2}\cdot S_{z4})} e^{2i\omega t_3 (S_{z3}\cdot S_{z4})} e^{-i\omega t_3 S_{z2}} e^{-i\omega t_3 S_{z3}} e^{i\omega\ {}_2 S_{y4}}$$

$$= e^{-i\frac{\pi}{4}\sigma_{y4}} U_\phi{}^{(1,4)}(t_1) e^{-i\frac{\pi}{16}(\sigma_{z1}\cdot\sigma_{z2})} e^{-i\frac{\pi}{16}(\sigma_{z1}\cdot\sigma_{z3})} e^{-i\frac{\pi}{16}(\sigma_{z2}\cdot\sigma_{z3})}$$

$$\times\, e^{-i\frac{\pi}{16}(\sigma_{z2}\cdot\sigma_{z4})} e^{-i\frac{\pi}{16}(\sigma_{z3}\cdot\sigma_{z4})} e^{i\frac{\pi}{16}\sigma_{z2}} e^{i\frac{\pi}{16}\sigma_{z3}} e^{i\frac{\pi}{4}\sigma_{y4}}$$

The argument on the left-hand side, $t_1 + 2\,t_2 + 7\,t_3$, indicates the total time taken by the various pulses, appearing on the right-hand side, in the execution of this gate.

- *CNOT gate, ($U^{(1,2)}{}_{CNOT}$), for four qubits where the first qubit is the control qubit and the second qubit is the target qubit, the third and fourth qubits are inactive*. Its quantum circuit is shown below

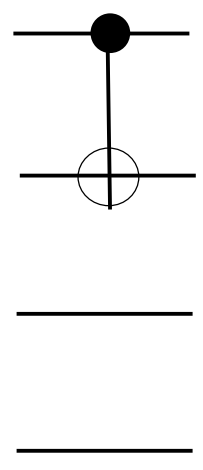

It is expressed as,

$$U^{(1,2)}{}_{CNOT}(T)$$

$$= e^{-i\omega t S_{y2}} U_\phi{}^{CNOT}(t) e^{2i\omega t (S_{z1}\cdot S_{z3})} e^{2i\omega t (S_{z1}\cdot S_{z4})} e^{2i\omega t (S_{z2}\cdot S_{z3})} e^{2i\omega t (S_{z2}\cdot S_{z4})} e^{2i\omega t (S_{z3}\cdot S_{z4})} e^{-i\omega t S_{z3}} e^{-i\omega t S_{z4}} e^{i\omega t S_{y2}}$$

where T is the total time taken by the pulses.

Now, if the value of $t_5$ is chosen such that

$$\frac{\omega t_5}{2} = 2m\pi + \frac{\pi}{4};\ m > 0,$$

where, m is an integer, one obtains

$$U^{(1,2)}{}_{CNOT}(t_4 + 9t_5) = e^{-i\omega t_5 S_{y2}} U_{\phi}{}^{CNOT}(t_4) e^{2i\omega t_5 (S_{z1} \cdot S_{z3})} e^{2i\omega t_5 (S_{z1} \cdot S_{z4})} e^{2i\omega t_5 (S_{z2} \cdot S_{z3})}$$
$$e^{2i\omega t_5 (S_{z2} \cdot S_{z4})} e^{2i\omega t_5 (S_{z3} \cdot S_{z4})} e^{-i\omega t_5 S_{z3}} e^{-i\omega t_5 S_{z4}} e^{i\omega t_5 S_{y2}}$$

$$= e^{-i\frac{\pi}{4}\sigma_{y2}} U_{\phi}{}^{CNOT}(t_8) e^{i\frac{\pi}{4}(\sigma_{z1} \cdot \sigma_{z3})} e^{i\frac{\pi}{4}(\sigma_{z1} \cdot \sigma_{z4})} e^{i\frac{\pi}{4}(\sigma_{z2} \cdot \sigma_{z3})} e^{i\frac{\pi}{4}(\sigma_{z2} \cdot \sigma_{z4})} e^{i\frac{\pi}{4}(\sigma_{z3} \cdot \sigma_{z4})} e^{-i\frac{\pi}{4}\sigma_{z3}} e^{-i\frac{\pi}{4}\sigma_{z4}} e^{i\frac{\pi}{4}\sigma_{y2}},$$

where $U_{\phi}{}^{CNOT}(t_4)$ is given by Equation (8). Here $T = t_4 + 9t_5$.

❖ *Controlled-$X^{\frac{1}{4}}$ gate for four qubits where the second qubit is the control qubit and the fourth qubit is the target qubit, first and third qubits are inactive* ($U^{(2,4)}{}_{SSX}$):

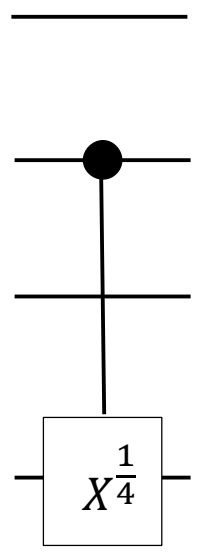

$U^{(2,4)}{}_{SSX}$ can be obtained by interchanging the first and the second qubits in $U^{(1,4)}{}_{SSX}$ with each other, except for keeping the matrix multiplication $(S_{z1} \cdot S_{z2})$ the same and changing $U_{\phi}{}^{(1,4)}(t_1)$ to $U_{\phi}{}^{(2,4)}(t_6)$. So, we obtain,

$$U^{(2,4)}{}_{SSX}(t_6 + 2\ t_7 + 7\ t_8) =$$
$$e^{-i\frac{\pi}{4}\sigma_{y4}} U_{\phi}{}^{(2,4)}(t_6) e^{-i\frac{\pi}{16}(\sigma_{z1} \cdot \sigma_{z2})} e^{-i\frac{\pi}{16}(\sigma_{z1} \cdot \sigma_{z3})} e^{-i\frac{\pi}{16}(\sigma_{z1} \cdot \sigma_{z4})} e^{-i\frac{\pi}{16}(\sigma_{z2} \cdot \sigma_{z3})} e^{-i\frac{\pi}{16}(\sigma_{z3} \cdot \sigma_{z4})} e^{i\frac{\pi}{16}\sigma_{z1}} e^{i\frac{\pi}{16}\sigma_{z3}} e^{i\frac{\pi}{4}\sigma_{y4}}$$

❖ *Controlled-$X^{-\frac{1}{4}}$ gate, $(U^{(2,4)}{}_{SSX})^{\dagger}$), for four qubits where the second qubit is the control qubit and the fourth qubit is the target qubit, first and third qubits are inactive.* The quantum circuit diagram for it is shown below.

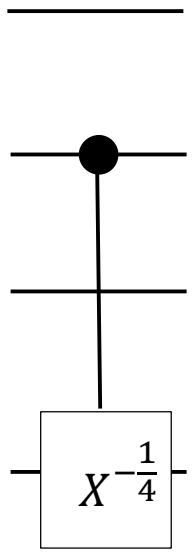

This gate is $(U^{(2,4)}{}_{SSX})^{\dagger}$ i.e., the adjoint of $U^{(2,4)}{}_{SSX}$, which has already been derived above.

❖ *CNOT gate for 4 qubits where the second qubit is the control qubit and the third qubit is the target qubit and the first and fourth qubits are inactive( $U^{(2,3)}{}_{CNOT}$):*

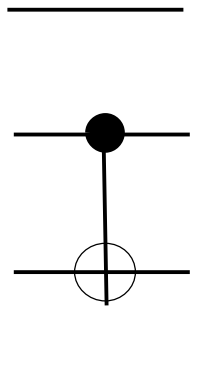

This CNOT gate is denoted by $U^{(2,3)}{}_{CNOT}$, where (2,3) denotes that the second qubit is the control qubit and the third qubit is the target qubit, to which the NOT gate will be applied, and nothing happens to the first and fourth qubit.

$$U^{(2,3)}{}_{CNOT}(T) = e^{-i\omega t S_{y3}} U_{\phi}{}^{CNOT}(t) e^{2i\omega t(S_{z1}\cdot S_{z2})} e^{2i\omega t(S_{z1}\cdot S_{z3})} e^{2i\omega t(S_{z1}\cdot S_{z4})} e^{2i\omega t(S_{z2}\cdot S_{z4})} e^{2i\omega t(S_{z3}\cdot S_{z4})} \times e^{-i\omega t S_{z1}} e^{-i\omega t S_{z4}} e^{i\omega t S_{y3}},$$

where T is the total time taken by the pulses.

The value of $t_{10}$ is chosen such that

$$\frac{\omega t_{10}}{2} = 2m\pi + \frac{\pi}{4};\ m > 0,$$

where, $m$ is an integer. then

$$U^{(2,3)}{}_{CNOT}(t_9 + 9t_{10}) = e^{-i\omega t_{10} S_{y3}} U_\phi{}^{CNOT}(t_9) e^{2i\omega t_{10}(S_{z1} \cdot S_{z2})} e^{2i\omega t_{10}(S_{z1} \cdot S_{z3})} e^{2i\omega t_{10}(S_{z1} \cdot S_{z4})}$$

$$\times e^{2i\omega t_{10}(S_{z2} \cdot S_{z4})} e^{2i\omega t_{10}(S_{z3} \cdot S_{z4})} e^{-i\omega t_{10} S_{z1}} e^{-i\omega t_{10} S_{z4}} e^{i\omega t_{10} S_{y3}}$$

$$= e^{-i\frac{\pi}{4}\sigma_{y3}} U_\phi{}^{CNOT}(t_9) e^{i\frac{\pi}{4}(\sigma_{z1} \cdot \sigma_{z2})} e^{i\frac{\pi}{4}(\sigma_{z1} \cdot \sigma_{z3})} e^{i\frac{\pi}{4}(\sigma_{z1} \cdot \sigma_{z4})}$$

$$\times e^{i\frac{\pi}{4}(\sigma_{z2} \cdot \sigma_{z4})} e^{i\frac{\pi}{4}(\sigma_{z3} \cdot \sigma_{z4})} e^{-i\frac{\pi}{4}\sigma_{z1}} e^{-i\frac{\pi}{4}\sigma_{z4}} e^{i\frac{\pi}{4}\sigma_{y3}}$$

- *Controlled-$X^{\frac{1}{4}}$ gate for 4 qubits where the third qubit is the control qubit and the fourth qubit is the target qubit and nothing happens to the first and second qubits ($U^{(3,4)}{}_{SSX}$):*

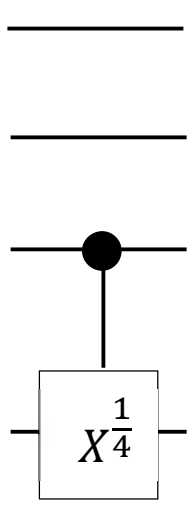

This controlled-$X^{\frac{1}{4}}$ 4-qubit gate is denoted as $U^{(3,4)}{}_{SSX}$, where (3,4) denotes the third qubit is the control qubit and the fourth qubit is the target qubit.

The operator $U^{(1,4)}{}_{SSX}$ can be obtained similarly by replacing qubit 1 with qubit 3 and qubit 3 with qubit 1, except for the matrix multiplication $(S_{z1} \cdot S_{z3})$, which is not changed. Finally, $U^{(3,4)}{}_{SSX}$ can be obtained from $U^{(1,4)}{}_{SSX}$ by corresponding changes in the positions of individual pulses and by taking the correct order of multiplication like $(S_{z2} \cdot S_{z3}), (S_{z1} \cdot S_{z2})$.

Therefore,

$$U^{(3,4)}{}_{SSX}(t_{11} + 2\, t_{12} + 7\, t_{13}) =$$

$$e^{-i\frac{\pi}{4}\sigma_{y4}} U_\phi{}^{(3,4)}(t_{11}) e^{-i\frac{\pi}{16}(\sigma_{z1} \cdot \sigma_{z2})} e^{-i\frac{\pi}{16}(\sigma_{z1} \cdot \sigma_{z3})} e^{-i\frac{\pi}{16}(\sigma_{z1} \cdot \sigma_{z4})} e^{-i\frac{\pi}{16}(\sigma_{z2} \cdot \sigma_{z3})} e^{-i\frac{\pi}{16}(\sigma_{z2} \cdot \sigma_{z4})} e^{i\frac{\pi}{16}\sigma_{z1}} e^{i\frac{\pi}{16}\sigma_{z2}} e^{i\frac{\pi}{4}\sigma_{y4}}$$

❖ *Controlled-X$^{-\frac{1}{4}}$ gate for 4 qubits where the 3rd qubit is the control qubit and the 4th qubit is the target qubit, doing nothing to the 1st and 2nd qubits* ($(U^{(3,4)}{}_{SSX})^{\dagger}$):

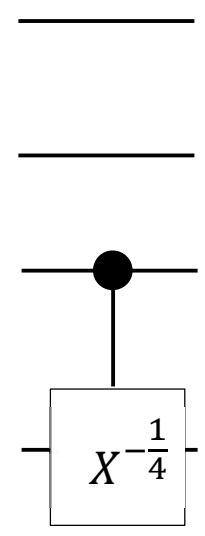

This gate is $(U^{(3,4)}{}_{SSX})^{\dagger}$, i.e., adjoint of $U^{(3,4)}{}_{SSX}$, which has already been discussed above.

❖ *CNOT gate for four qubits, where the first qubit is the control qubit and the third qubit is the target qubit,* to which the NOT gate will be applied, nothing happens to the second and fourth qubits ($U^{(1,3)}{}_{CNOT}$):

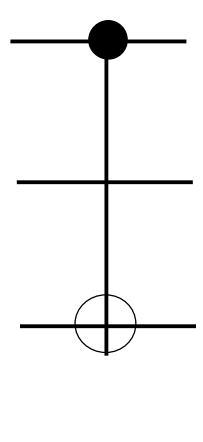

This operator can be obtained similarly to $U^{(2,3)}{}_{CNOT}$ by interchanging qubit 2 and qubit 1 with each other, except for retaining the matrix multiplication $(S_{z1} \cdot S_{z2})$ to be the same. Finally, $U^{(1,3)}{}_{CNOT}$ is obtained by making appropriate changes in the positions of the individual pulses.

Therefore,

$$U^{(1,3)}{}_{CNOT} \quad (t_{14} + 9t_{15})$$
$$= e^{-i\frac{\pi}{4}\sigma_{y3}} U_{\phi}{}^{CNOT}(t_{14}) e^{i\frac{\pi}{4}(\sigma_{z1}\cdot\sigma_{z2})} e^{i\frac{\pi}{4}(\sigma_{z1}\cdot\sigma_{z4})} e^{i\frac{\pi}{4}(\sigma_{z2}\cdot\sigma_{z3})} e^{i\frac{\pi}{4}(\sigma_{z2}\cdot\sigma_{z4})} e^{i\frac{\pi}{4}(\sigma_{z3}\cdot\sigma_{z4})} e^{-i\frac{\pi}{4}\sigma_{z2}} e^{-i\frac{\pi}{4}\sigma_{z4}} e^{i\frac{\pi}{4}\sigma_{y3}}$$

**Construction of CCCNOT gate**

In the quantum circuit shown in Figure 4, the component gates are from left to right as follows:

$U^{(1,4)}{}_{SSX}$, $U^{(1,2)}{}_{CNOT}$, $U^{(2,4)}{}_{SSX})^{\dagger}$ , $U^{(1,2)}{}_{CNOT}$, $U^{(2,4)}{}_{SSX}$, $U^{(2,3)}{}_{CNOT}$, $U^{(3,4)}{}_{SSX})^{\dagger}$ , $U^{(1,3)}{}_{CNOT}$ , $U^{(3,4)}{}_{SSX}$ , $U^{(2,3)}{}_{CNOT}$ , $U^{(3,4)}{}_{SSX})^{\dagger}$ , $U^{(1,3)}{}_{CNOT}$ , $U^{(3,4)}{}_{SSX}$

For the various time intervals required here, the following notations are used for convenience:

$$T_1 = t_1 + 2\,t_2 + 7\,t_3$$

$$T_2 = t_4 + 9t_5$$

$$T_3 = t_6 + 2\,t_7 + 7\,t_8$$

$$T_4 = t_9 + 9t_{10}$$

$$T_5 = t_{11} + 2\,t_{12} + 7\,t_{13}$$

$$T_6 = t_{14} + 9t_{15}$$

$$T = T_1 + 2T_2 + 2T_3 + 2T_4 + 4T_5 + 2T_6$$

To obtain the final 4-qubit gate producing the $16 \times 16$ Toffoli matrix, the various component gates are multiplied in the sequence from left to right, specified above:

$$\begin{aligned} U_{CCCNOT}(T) = {} & U^{(3,4)}{}_{SSX}(T_5) \cdot U^{(1,3)}{}_{CNOT}(T_6) \cdot (U^{(3,4)}{}_{SSX})^{\dagger}(T_5) \cdot U^{(2,3)}{}_{CNOT}(T_4) \cdot \\ & \cdot U^{(3,4)}{}_{SSX}(T_5) \cdot U^{(1,3)}{}_{CNOT}(T_6) \cdot (U^{(3,4)}{}_{SSX})^{\dagger}(T_5) \cdot U^{(2,3)}{}_{CNOT}(T_4) \cdot \\ & \cdot U^{(2,4)}{}_{SSX}(T_3) \cdot U^{(1,2)}{}_{CNOT}(T_2) \cdot (U^{(2,4)}{}_{SSX})^{\dagger}(T_3) \cdot U^{(1,2)}{}_{CNOT}(T_2) \cdot \\ & \cdot U^{(1,4)}{}_{SSX}(T_1) \end{aligned}$$

$$
= \begin{bmatrix}
1 & 0 & 0 & 0 & 0 & 0 & 0 & 0 & 0 & 0 & 0 & 0 & 0 & 0 & 0 & 0 \\
0 & 1 & 0 & 0 & 0 & 0 & 0 & 0 & 0 & 0 & 0 & 0 & 0 & 0 & 0 & 0 \\
0 & 0 & 1 & 0 & 0 & 0 & 0 & 0 & 0 & 0 & 0 & 0 & 0 & 0 & 0 & 0 \\
0 & 0 & 0 & 1 & 0 & 0 & 0 & 0 & 0 & 0 & 0 & 0 & 0 & 0 & 0 & 0 \\
0 & 0 & 0 & 0 & 1 & 0 & 0 & 0 & 0 & 0 & 0 & 0 & 0 & 0 & 0 & 0 \\
0 & 0 & 0 & 0 & 0 & 1 & 0 & 0 & 0 & 0 & 0 & 0 & 0 & 0 & 0 & 0 \\
0 & 0 & 0 & 0 & 0 & 0 & 1 & 0 & 0 & 0 & 0 & 0 & 0 & 0 & 0 & 0 \\
0 & 0 & 0 & 0 & 0 & 0 & 0 & 1 & 0 & 0 & 0 & 0 & 0 & 0 & 0 & 0 \\
0 & 0 & 0 & 0 & 0 & 0 & 0 & 0 & 1 & 0 & 0 & 0 & 0 & 0 & 0 & 0 \\
0 & 0 & 0 & 0 & 0 & 0 & 0 & 0 & 0 & 1 & 0 & 0 & 0 & 0 & 0 & 0 \\
0 & 0 & 0 & 0 & 0 & 0 & 0 & 0 & 0 & 0 & 1 & 0 & 0 & 0 & 0 & 0 \\
0 & 0 & 0 & 0 & 0 & 0 & 0 & 0 & 0 & 0 & 0 & 1 & 0 & 0 & 0 & 0 \\
0 & 0 & 0 & 0 & 0 & 0 & 0 & 0 & 0 & 0 & 0 & 0 & 1 & 0 & 0 & 0 \\
0 & 0 & 0 & 0 & 0 & 0 & 0 & 0 & 0 & 0 & 0 & 0 & 0 & 1 & 0 & 0 \\
0 & 0 & 0 & 0 & 0 & 0 & 0 & 0 & 0 & 0 & 0 & 0 & 0 & 0 & 0 & 1 \\
0 & 0 & 0 & 0 & 0 & 0 & 0 & 0 & 0 & 0 & 0 & 0 & 0 & 0 & 1 & 0
\end{bmatrix},
$$

which is the four-qubit $16 \times 16$ Toffoli matrix.

## 4. Conclusions

The theoretical details of how to construct EPR-based Toffoli gates for n-qubits (n=1,2,3,4), presented in this paper, are *avant-garde*. When the experimental EPR pulse technology is developed sufficiently to realize the proposed experiments, involving resonance and application of pulses, it will be possible to build EPR-based quantum computers using these gates. This is quite feasible in the near future, as experimental technology is evolving very rapidly.

It is noted that NMR cannot be used for this purpose as nuclear spins are not amenable to coupling by the exchange interaction, required in the proposed construction.

***Acknowledgments:*** We are grateful to MITACS for providing funding to support the sojourn of Sayan Manna during summer 2024 at Concordia University.

***Code Availability:*** The code used to generate the results of this study is publicly available at: [https://github.com/sayanmanna1/Toffoli-gate-Simulation]. All simulations and analysis scripts are included.

## Appendix I

### Fault-tolerant quantum computation

The dawn of fault-tolerant quantum computing (FTQC) will signal the ability of quantum computers to perform calculations with arbitrarily low logical error rates. The qubit counts of today are, in principle, high enough to solve certain real-world problems. Unfortunately, the error rates are still too high to make the results of any computation useful. FTQC will signal:

- The error correction threshold has been exceeded.
- Quantum information is adequately shielded from the environment.
- The spread of errors throughout the array is contained locally.

## Appendix II

### Solution of Schrödinger equation in the rotating frame

In the EPR experiment, there is a static, time-independent, magnetic field in the $z$ direction, $\overrightarrow{B_0}$, and a circularly polarized magnetic field, $\overrightarrow{B_1}$, rotating in the $xy$ plane. Thus, the total magnetic field is,

$$\vec{B}(t) = B_0\hat{e}_z + B_1(\hat{e}_x \cos(\omega t) - \hat{e}_y \sin(\omega t))$$

Typically, the constant $z$-component $B_0 >> B_1$ , the magnitude of the RF (radio- frequency) signal. The time dependent part of the magnetic field points along the $x$-axis at $t = 0$ and is rotating with angular velocity ω in the clockwise direction of the $xy$ plane. The spin Hamiltonian is

$$H_s = -\gamma\, \vec{S} \cdot \vec{B}$$

$$= -\gamma[B_0 S_z + B_1(S_x \cos(\omega t) - S_y \sin(\omega t))]$$

Let the initial state be $\psi(0) = a_0|\uparrow\rangle + b_0|\downarrow\rangle = \begin{bmatrix} a_0 \\ b_0 \end{bmatrix}$

Then, the state vector at time t is denoted by

$$\psi(t) = a(t)|\uparrow\rangle + b(t)|\downarrow\rangle = \begin{bmatrix} a(t) \\ b(t) \end{bmatrix},$$

where $a(t), b(t) \in \mathbb{C}$ and satisfy $|a(t)|^2 + |b(t)|^2 = 1$

Now the Schrödinger equation is

$$i\hbar\frac{\partial}{\partial t}\psi(t) = \hat{H}\psi(t),$$

$$i\hbar\frac{\partial}{\partial t}\psi(t) = -\gamma[B_0 S_z + B_1(S_x \cos(\omega t) - S_y \sin(\omega t))]\psi(t)$$

Here the Hamiltonian is time dependent; as well the Hamiltonian at different times do not commute. To solve this problem, a different frame of reference is considered that rotates with the magnetic field and that frame of reference rotates about the $z$-axis with a frequency $\omega$. It will be denoted by $R$. In this frame the spin states that are fixed in the original frame would be seen to rotate with positive angular velocity $\omega$ about the $z$ direction. There must be a Hamiltonian that does have that effect. So, the Hamiltonian for this rotating frame of reference is, say $H_U$. So, solving the Schrödinger equation for this case one gets the time evolution unitary operator $U$ which does this rotation:

$$U(t) = e^{-\frac{i\omega t S_z}{\hbar}} \Rightarrow H_U = \omega S_z$$

So, this Hamiltonian $H_U$ is time independent.

For the original case, the Hamiltonian in the static frame is $H_S$, which is time dependent.

The above operator $U(t)$ is used to define a new state in the rotating-frame, $|\psi_R\rangle$, as follows:

$$|\psi_R, t\rangle = U(t)|\psi, t\rangle,$$

where $|\psi_R, t\rangle$ is the state in the rotating frame of reference at time t and $|\psi, t\rangle$ is the original state at time t.

So, $|\psi, 0\rangle$ is the original state at $t = 0$, then upon applying the total magnetic field it will evolve according to the original Hamiltonian $H_S$, with the corresponding operator being $U_S$. Then perceiving the quantum state from the rotation frame R, where the original state is evolving according to the Hamiltonian $H_U$ with the corresponding unitary operator $U$.

$$|\psi, t\rangle = U_S|\psi, 0\rangle$$

Therefore,

$$|\psi_R, t\rangle = U(t)U_S(t)|\psi, t\rangle$$

Now, one solves the Schrödinger equation for $|\psi_R\rangle$, which is simpler. If one knows $|\psi_R\rangle$, then one can get the original state $|\psi\rangle$. The corresponding Hamiltonian in the rotating frame of reference is $H_R$. It will now be shown that this Hamiltonian $H_R$ is time independent.

Since the Hamiltonian associated to an arbitrary unitary time evolution operator $U$ is $i\hbar(\partial_t U)U^\dagger$. Then,

$$\begin{aligned} H_R &= i\hbar\partial_t(UU_S){U_S}^\dagger U^\dagger \\ &= i\hbar(\partial_t U)U^\dagger + Ui\hbar(\partial_t U_S){U_S}^\dagger U^\dagger \\ &= H_U + UH_SU^\dagger, \end{aligned}$$

where $H_U = i\hbar(\partial_t U)U^\dagger$ and one knows that $H_U = \omega S_z$.

So,

$$H_R = \omega S_z + e^{-\frac{i\omega t S_z}{\hbar}}[-\gamma(B_0S_z + B_1\left(S_x\cos{(\omega t)} - S_y\sin{(\omega t)}\right))]e^{\frac{i\omega t S_z}{\hbar}}$$

$$= (-\gamma B_0 + \omega)S_z - \gamma B_1 e^{-\frac{i\omega t S_z}{\hbar}} \left(S_x \cos(\omega t) - S_y \sin(\omega t)\right) e^{\frac{i\omega t S_z}{\hbar}}$$

Let,

$$M(t) = e^{-\frac{i\omega t S_z}{\hbar}} \left(S_x \cos(\omega t) - S_y \sin(\omega t)\right) e^{\frac{i\omega t S_z}{\hbar}}$$

Analyzing $M(t)$, if one takes the first order time derivative of $M$, one finds

$$\partial_t M = 0$$

So, there is no time dependence of $M$. This indicates that $H_R$ is time independent Hamiltonian. Then, one can choose at any time $t = t$, in particular, choose $t = 0$, to find that

$$M = S_x$$

Therefore,

$$H_R = (-\gamma B_0 + \omega)S_z - \gamma B_1 S_x$$

So, the Schrödinger equation is

$$i\hbar \frac{\partial}{\partial t} \psi_R = H_R \psi_R$$

$$\Rightarrow |\psi_R, t\rangle = exp\,[-\frac{iH_R t}{\hbar}]|\psi_R, 0\rangle$$

$$\Rightarrow |\psi_R, t\rangle = exp\,[-\frac{i((-\gamma B_0 + \omega)S_z - \gamma B_1 S_x)t}{\hbar}]|\psi_R, 0\rangle$$

Now, the full solution of the original Schrödinger equation is

$$|\psi, t\rangle = U^\dagger(t)|\psi_R, t\rangle = exp\,[\frac{i\omega t S_z}{\hbar}]|\psi_R, t\rangle$$

Since $H_R$ is a time-independent Hamiltonian, the full-time evolution is given by

$$|\psi, t\rangle = exp\,[\frac{i\omega t S_z}{\hbar}]exp\,[-\frac{iH_R t}{\hbar}]|\psi, 0\rangle$$

$$= exp\,[\frac{i\omega t S_z}{\hbar}]exp\,[-\frac{i((-\gamma B_0 + \omega)S_z - \gamma B_1 S_x)t}{\hbar}]|\psi, 0\rangle$$

This is the final solution of the non-static magnetic field for one particle.

## Appendix III

**Flowchart for simulating n-qubits (n=2,3,4) Toffoli gates in Python**

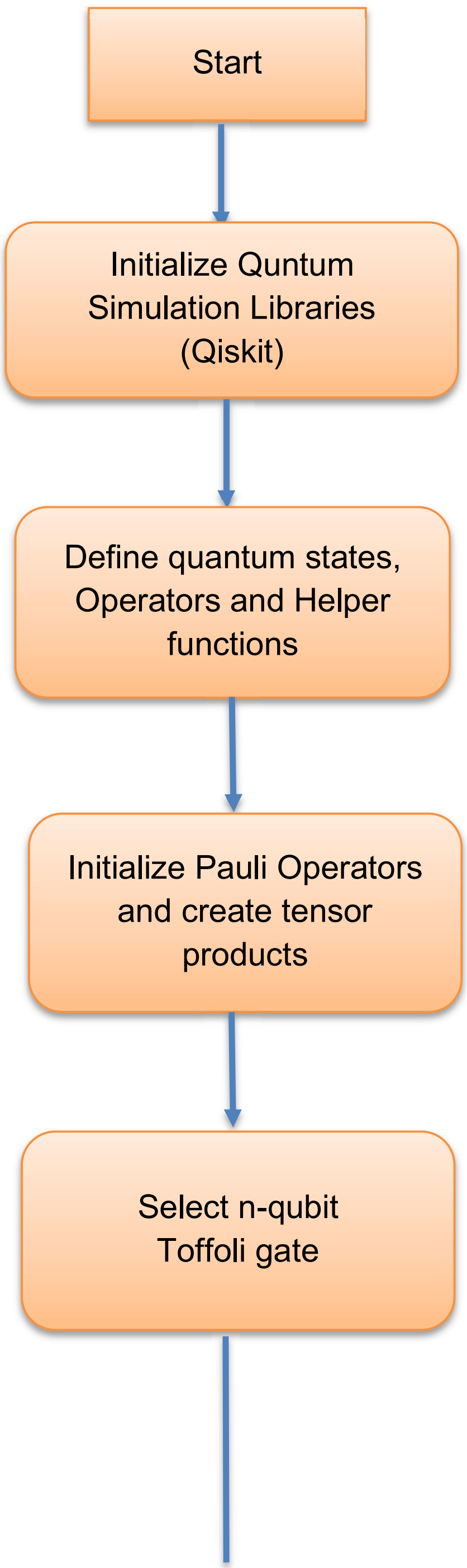

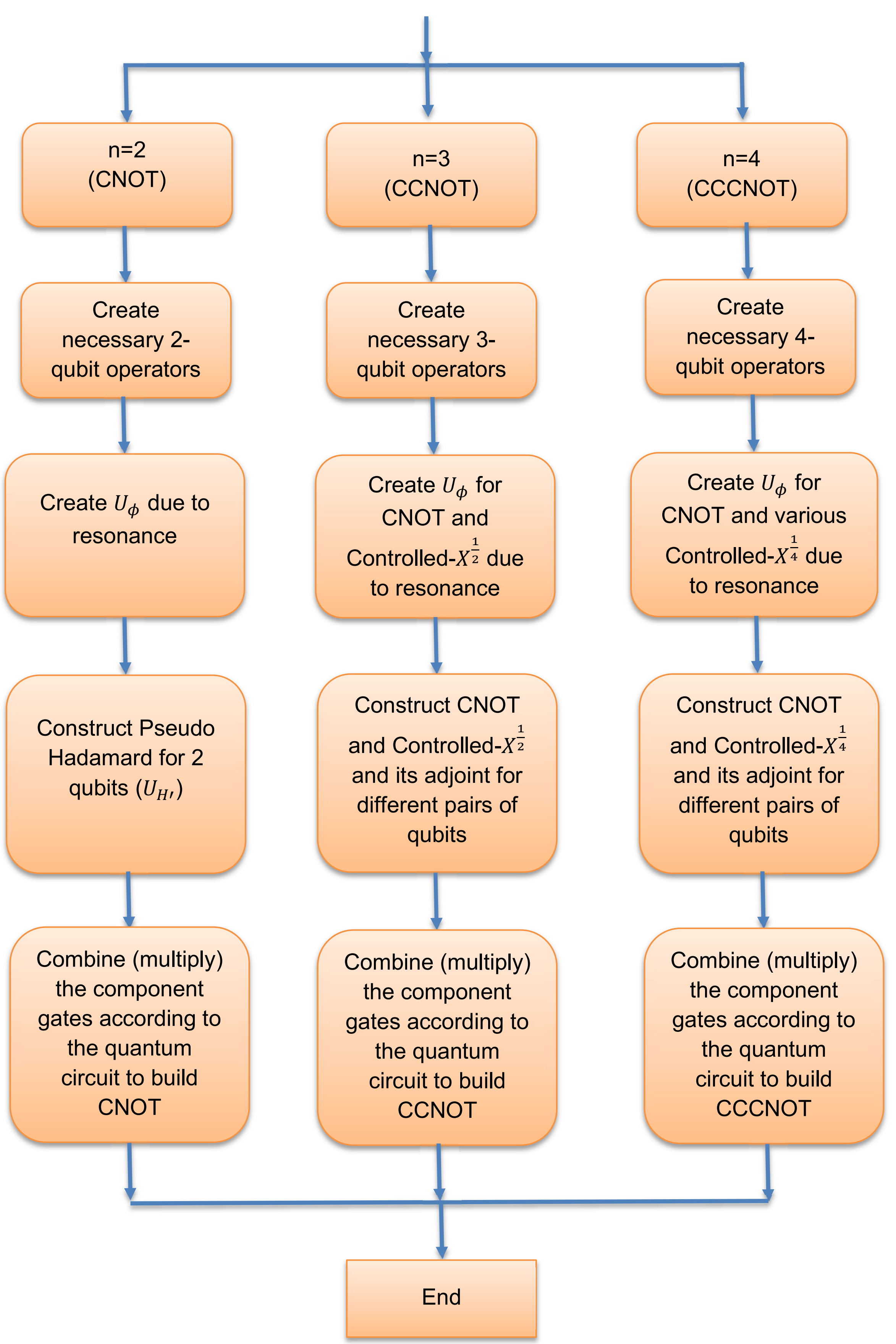

n=2 (CNOT)
n=3 (CCNOT)
n=4 (CCCNOT)
Create necessary 2-qubit operators
Create necessary 3-qubit operators
Create necessary 4-qubit operators
Create $U_\phi$ due to resonance
Create $U_\phi$ for CNOT and Controlled-$X^{\frac{1}{2}}$ due to resonance
Create $U_\phi$ for CNOT and various Controlled-$X^{\frac{1}{4}}$ due to resonance
Construct Pseudo Hadamard for 2 qubits ($U_{H'}$)
Construct CNOT and Controlled-$X^{\frac{1}{2}}$ and its adjoint for different pairs of qubits
Construct CNOT and Controlled-$X^{\frac{1}{4}}$ and its adjoint for different pairs of qubits
Combine (multiply) the component gates according to the quantum circuit to build CNOT
Combine (multiply) the component gates according to the quantum circuit to build CCNOT
Combine (multiply) the component gates according to the quantum circuit to build CCCNOT
End